\documentclass[showkeys,groupedaddress,nofootinbib,showpacs,eqsecnum]{revtex4}

\usepackage{euscript,graphicx,amssymb,subfigure}

\usepackage{amsfonts}
\usepackage{amsmath}
\usepackage{amssymb}
\usepackage{graphicx}
\usepackage{euscript}
\usepackage{color}
\usepackage{subfigure}

\begin{document}

\title{Chameleon gravity on cosmological scales  }

\author{H. Farajollahi$^{1,2}$}
\email{hosseinf@guilan.ac.ir} \author{A. Salehi$^{1}$}
\affiliation{$^1$Department of Physics, University of Guilan, Rasht, Iran}
\affiliation{$^2$ School of Physics, University of New South Wales, Sydney, NSW, 2052, Australia}

\begin{abstract}

In conventional approach to the chameleon mechanism, by assuming a static and spherically symmetric solutions in which matter density and chameleon field are given by $\rho=\rho(r)$ and $\phi=\phi(r)$, it has been shown that mass of chameleon field is matter density-dependent. In regions of high matter density such as earth, chameleon field is massive, in solar system it is low and in cosmological scales it is very low. In this article we revisit the mechanism in cosmological scales by assuming a redshift dependence of the matter density and chameleon field, i.e. $\rho=\rho(z)$, $\phi=\phi(z)$. To support our analysis, we best fit the model parameters with the observational data. The result shows that in cosmological scales, the mass of chameleon field increases with the redshift, i.e. more massive in higher redshifts. We also find that in both cases of power-law and exponential potential function, the current universe acceleration can be explained by the low mass chameleon field. In comparison with the high redshift observational data, we also find that the model with power-law potential function is in better agreement with the observational data.

\end{abstract}


\maketitle

\section{Introduction}\label{s:intro}

Recently, the observations of high redshift type Ia
supernovae and the surveys of clusters of galaxies \cite{Reiss}--\cite{Pope} reveal cosmic acceleration. Further,
observations of Cosmic Microwave Background (CMB)
anisotropies indicates that the universe is flat and the total energy
density is very close to the critical one \cite{Spergel}.

These data properly complete each other and designate that the so called dark
energy (DE) is the dominant component of the present universe, occupies about $\%73$ of the cosmic energy , while dark matter (DM) occupies $\%23$, and the usual
baryonic matter takes about $\%4$. There are prominent candidates for DE such as the cosmological
constant \cite{Sahni, Weinberg}, a dynamically evolving scalar field ( like quintessence) \cite{Caldwell, Zlatev} or phantom field ( with negative energy) \cite{Caldwell2} that explain cosmic accelerating expansion. Alternatively, the universe acceleration can also be elucidated through
modified gravity \cite{Zhu},  brane cosmology and so on \cite{Zhu1}--\cite{set10}. The DE
can track the evolution of the background matter in the
early stage, and only recently, it becomes dominant . Thus, its current
condition is nearly independent of the initial conditions \cite{Lyth}--\cite{Easson}.

On the other hand, to explain the early and late time acceleration of the
universe. it is most often the case that such fields interact with matter; directly through matter Lagrangian
coupling, or indirectly by coupling to the Ricci scalar as emerged from quantum loop corrections \cite{Damouri}--\cite{Biswass}. If the
scalar field self-interactions are negligible, then the experimental bounds on such a field are very strong; requiring it to either
couple to matter much more weakly than gravity does, or to be very heavy \cite{Uzan}--\cite{Damourm}. Unfortunately, such fields are usually very light and its coupling to matter should be tuned to
extremely small values in order not to be conflict with the Equivalence Principal \cite{nojiri}.

An attempt to overcome the problem with light scalar fields has
been suggested in chameleon cosmology \cite{Khoury}--\cite{Khourym}. In
the proposed  model, a scalar field couples to matter with gravitational strength, in harmony with general expectations from
string theory whilst at the same time remaining very light on cosmological scales. The light field on cosmological scales is permitted
to couple to matter much more strongly than gravity does, and yet still satisfies the current experimental constraints. The cosmological value of such a field evolves over Hubble time-scales and could potentially cause the late-time acceleration of the Universe \cite{Brax2}. In this approach the mass of the scalar field depends on the
local background matter density. While the idea of a density-dependent mass term is not new \cite{Wett}--\cite{Mot}, in the work presented in \cite{Khourym} \cite{Brax2} the scalar field couples directly to matter with gravitational strength. In this paper, we stat with the assumption that the matter energy density is redshift dependent, $\rho=\rho(z)$. We therefore find that the chameleon scalar field has a greater contribution to the dynamic of the universe in higher redshifts.

The manuscript is organized as follows. In section two we describe the chameleon mechanism. We assume two forms of power law and exponential potential functions and derive the required equations. Section three is devoted to constrain the model parameters with the observational data using $\chi^2$ statistical method. In section four with numerical calculations and best fitted model parameters, we revisit the chameleon mechanism and test the model against recent findings. Summary and remarks are given in section five.

\section{The Model}

We begin with the action of chameleon gravity given by,
\begin{eqnarray}\label{action}
S=\int[\frac{M_{Pl}^2}{16\pi}{\cal R}-\frac{1}{2}\phi_{,\mu}\phi^{,\mu}+V(\phi)]\sqrt{-g}dx^{4}\nonumber\\+\int {\cal L}_m(\psi^{(i)}, g_{\mu\nu}^{(i)})dx^{4},
\end{eqnarray}
where the matter fields $\psi^{(i)}$ are coupled to scalar field $\phi$ by the definition $g_{\mu\nu}^{(i)}\equiv e^{2\beta_i\phi/M_{Pl}}g_{\mu\nu}$.
The $\beta_{i}$ are dimensionless coupling constants, one for each matter species. In the following, we assume a single matter energy density component $\rho_m$ with coupling $\beta$ \cite{Khourym}. The variation of action (\ref{action})  with respect to the metric tensor components in a spatially flat FRW  cosmology yields the field equations,
\begin{eqnarray}\label{fried1}
3H^{2}M_{pl}^{2}=\rho_{m}e^{\frac{\beta}{M_{pl}}\phi}+\frac{1}{2}\dot{\phi}^{2}+V(\phi),
\end{eqnarray}
\begin{eqnarray}\label{fried2}
(2\dot{H}+3H^2)M_{pl}^{2}=-\gamma \rho_{m}e^{\frac{\beta}{M_{pl}}\phi}-\frac{1}{2}\dot{\phi}^{2}+V(\phi).
\end{eqnarray}
In deriving the field equations we assumed a perfect fluid for matter field with $p_{m}=\gamma\rho_{m}$.
Variation of the action (\ref{action}) with respect to scalar field  $\phi$ provides the wave
equation for chameleon field as
\begin{eqnarray}\label{phiequation}
\ddot{\phi}+3H\dot{\phi}=-V^{'}-\frac{\beta}{M_{pl}}\rho_{m}e^{\frac{\beta}{M_{pl}}\phi},
\end{eqnarray}
where prime indicated differentiation with respect to $\phi$.
From equations (\ref{fried1}), (\ref{fried2}) and (\ref{phiequation}), one can easily arrive at the  relation
\begin{eqnarray}\label{conserv}
\dot{\rho_{m}}+3H\rho_{m}(1+\gamma)=-3\gamma\frac{\beta}{M_{pl}}\rho_{m}\dot{\phi}.
\end{eqnarray}
Integrating the above equation yields
\begin{eqnarray}\label{rom}
\rho_{m}=\frac{A}{e^{\frac{3\gamma\beta}{M_{pl}}\phi}a^{3(1+\gamma)}},
\end{eqnarray}
where  $A$  as a constant of integration. In the following we discuss two power-law and exponential forms for the potential in the model.

{\bf Case 1:}

The runaway inverse
power-law potential, also called the Ratra-Peebles potential, in chameleon cosmology is often given as
\begin{eqnarray}\label{veff00}
V(\phi)=\frac{M^{4+\alpha}}{\phi^{\alpha}},
\end{eqnarray}
where $M$ is a constant with the dimension of mass and $\alpha$ is a positive constant. These kind of potentials are usually seen in quintessence models. The chameleon effective potential is then defined by,
\begin{eqnarray}\label{veff0}
V_{eff}(\phi)=V(\phi)+\rho_{m}e^{\frac{\beta}{M_{pl}}\phi},
\end{eqnarray}
the sum of potential (\ref{veff00}) and its coupling to the matter density.
Using equation (\ref{rom}), the effective potential can be rewritten as
\begin{eqnarray}\label{veff20}
V_{eff}(\phi)=\frac{M^{4+\alpha}}{\phi^{\alpha}}+Ae^{\frac{(1-3\gamma)\beta}{M_{pl}}\phi}a^{-3(1+\gamma)}.
\end{eqnarray}
One can also easily find the mass associated with the field $\phi$ from:
\begin{eqnarray}
m^{2}=\frac{d^{2}}{d\phi^{2}}V_{eff}(\phi).
\end{eqnarray}
For matter dominated universe, $\gamma=0$, and positive $\beta$ in the second term of $V_{eff}(\phi)$, the effective potential monotonically decreases to a minimum at a finite field value $\phi=\phi_{min}$, where $\frac{d}{d\phi}V_{eff}|_{\phi=\phi_{min}}=0$, and $ m=m_{min}$. We then lead to the following relation
\begin{eqnarray}\label{veff40}
\phi_{min}^{\alpha+1}e^{\lambda\phi_{min}} = \frac{d}{\lambda},
\end{eqnarray}
where $\lambda=(1-3\gamma)\frac{\beta}{M_{pl}}$ and $d=\frac{\alpha M^{4+\alpha}a^{3(1+\gamma)}}{A}$. From equation (\ref{veff40}) we find
\begin{eqnarray}\label{phimin0}
\phi_{min} = \frac{1}{\lambda} W(\lambda(\frac{d}{\lambda})^{\frac{1}{1+\alpha}}),
\end{eqnarray}
where $W$ is called $Lambert$ or product-log function.
We can also find minimum mass from
\begin{eqnarray}\label{min0}
m_{min}^{2} = \alpha(\alpha+1)\frac{M^{4+\alpha}}{\phi_{min}^{\alpha+2}}+A\lambda^{2}e^{\lambda\phi_{min}}a^{-3(1+\gamma)}.
\end{eqnarray}
Using equation (\ref{veff40}) we can rewrite equation (\ref{min0}) as
\begin{eqnarray}\label{min2}
m_{min}^{2} =\alpha M^{4+\alpha}(\frac{\alpha+1}{\phi_{min}^{\alpha+2}}+\frac{\lambda}{\phi_{min}^{\alpha+1}}).
\end{eqnarray}
 From $a=\frac{a_{0}}{1+z}$ we find $d=\frac{d_{0}}{(1+z)^{3(1+\gamma)}}$ where $d_{0}=\frac{\alpha M^{4+\alpha}a_{0}^{3(1+\gamma)}}{A}$ is a positive constant. Thus, $\phi_{min}$ can be rewritten as
\begin{eqnarray}\label{phimin}
\phi_{min} = \frac{1}{\lambda} W(\lambda[\frac{d_{0}}{\lambda}(1+z)^{-3(1+\gamma)}]^{\frac{1}{1+\alpha}})
\end{eqnarray}

{\bf Case2:}

The exponential potential given by
\begin{eqnarray}
V(\phi)=M^{4}e^{(\frac{M^{\alpha}}{\phi^{\alpha}})},
\end{eqnarray}
is also seen in many cosmological models. Similar to the previous case, we find the effective potential, $V_{eff}(\phi)$, the minimum mass $m_{min}$, and the minimum scalar field, $\phi_{min}$, respectively as
\begin{eqnarray}\label{veff2}
V_{eff}(\phi)=M^{4}e^{(\frac{M^{\alpha}}{\phi^{\alpha}})}+Ae^{\frac{(1-3\gamma)\beta}{M_{pl}}\phi}a^{-3(1+\gamma)}
\end{eqnarray}
\begin{eqnarray}\label{min3}
m_{min}^{2} =\alpha M^{4+\alpha}(\frac{\alpha+1}{\phi_{min}^{\alpha+2}}+\frac{\lambda}{\phi_{min}^{\alpha+1}})
\end{eqnarray}
and
\begin{eqnarray}\label{phimin2}
\phi_{min} = \frac{1}{\lambda} W (\lambda[\frac{d_{0}}{\lambda}(1+z)^{-3(1+\gamma)}]^{\frac{1}{1+\alpha}}).
\end{eqnarray}
Note that $m_{min}$ is the inverse of the characteristic range of
the chameleon force in a given medium. In the next section we best fit the model with the observational data.

\section{Observational constraints}

We constrain the dimensionless coupling constant, $\beta$, and model parameter $\alpha$, in both exponential and power law potential, with the most recent observational data, SNe Ia, by employing the $\chi^2$ statistics. We assume that the energy density, $\rho_{m}$, stands for the contribution
from cold dark matter, $\gamma=0$. We also assume that $M_{pl}=1$, $M=1$, $A=1$. From Table \ref{table:1} one finds the best fitted model parameters $\alpha$ and  $\beta$ in both cases of power law and exponential potentials.

\begin{table}[ht]
\caption{best fitted parameters }  
\centering 
\begin{tabular}{c|c|c|c} 
\hline\hline 
Model  &  $\alpha$ \ & $\beta$ \ & $\chi^2$ \\
\hline 
power law & 0.175 & 4.51 & 557.0747981 \\ 
\hline 
exponential & 0.51 & 2.31 & 557.76250 \\
\end{tabular}
\label{table:1} 
\end{table}\

The confidence levels corresponding to the best fitted parameters are shown in FIG. 1). One sees that with $68.3\%$, $95.4\%$ and $99.7\%$ confidence level the true values for both $\alpha$ ,$\beta$ lie within the red, green and blue contours, respectively.

 \begin{figure}[t]
\includegraphics[scale=.25]{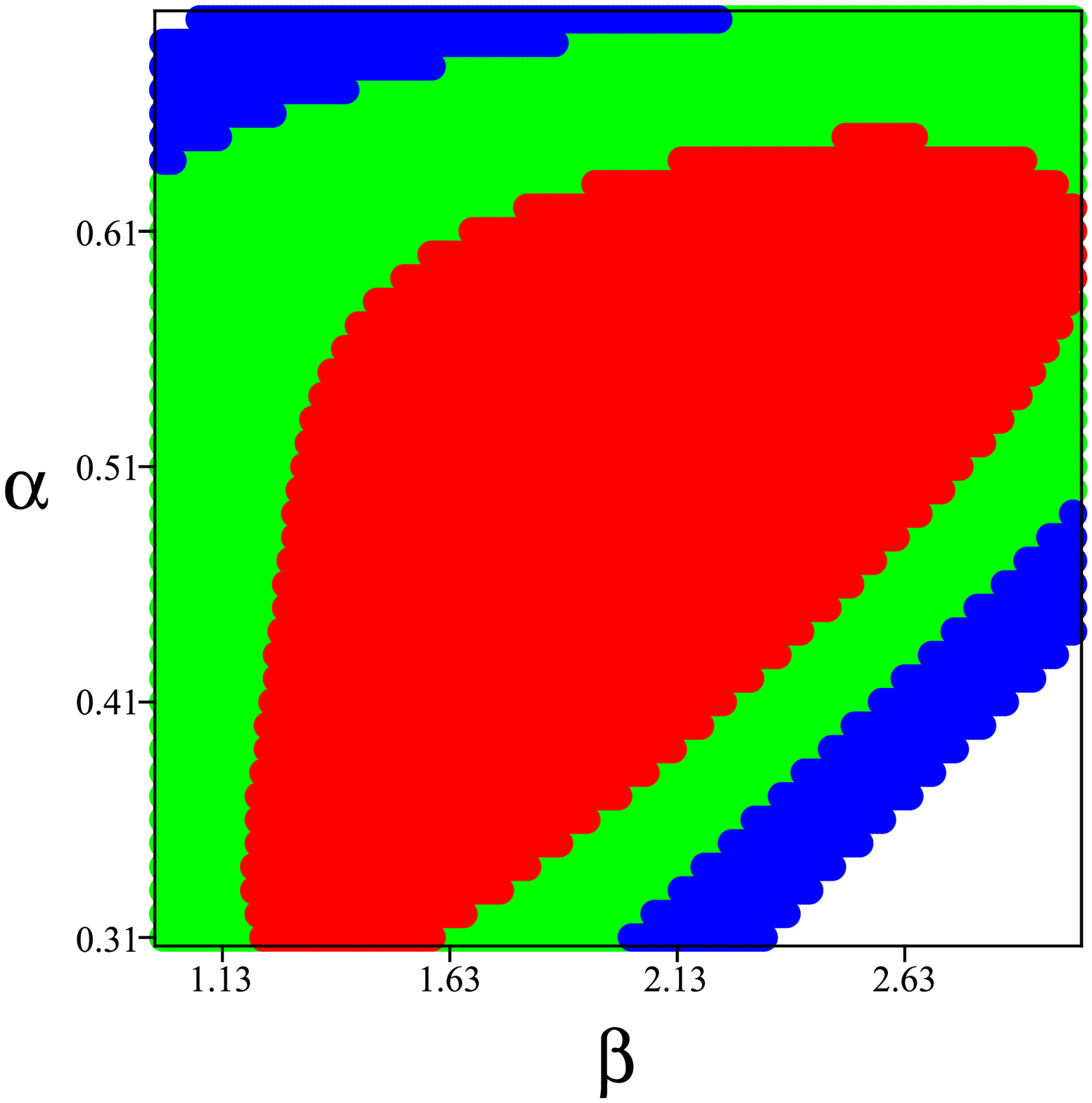}\hspace{0.1 cm}\includegraphics[scale=.25]{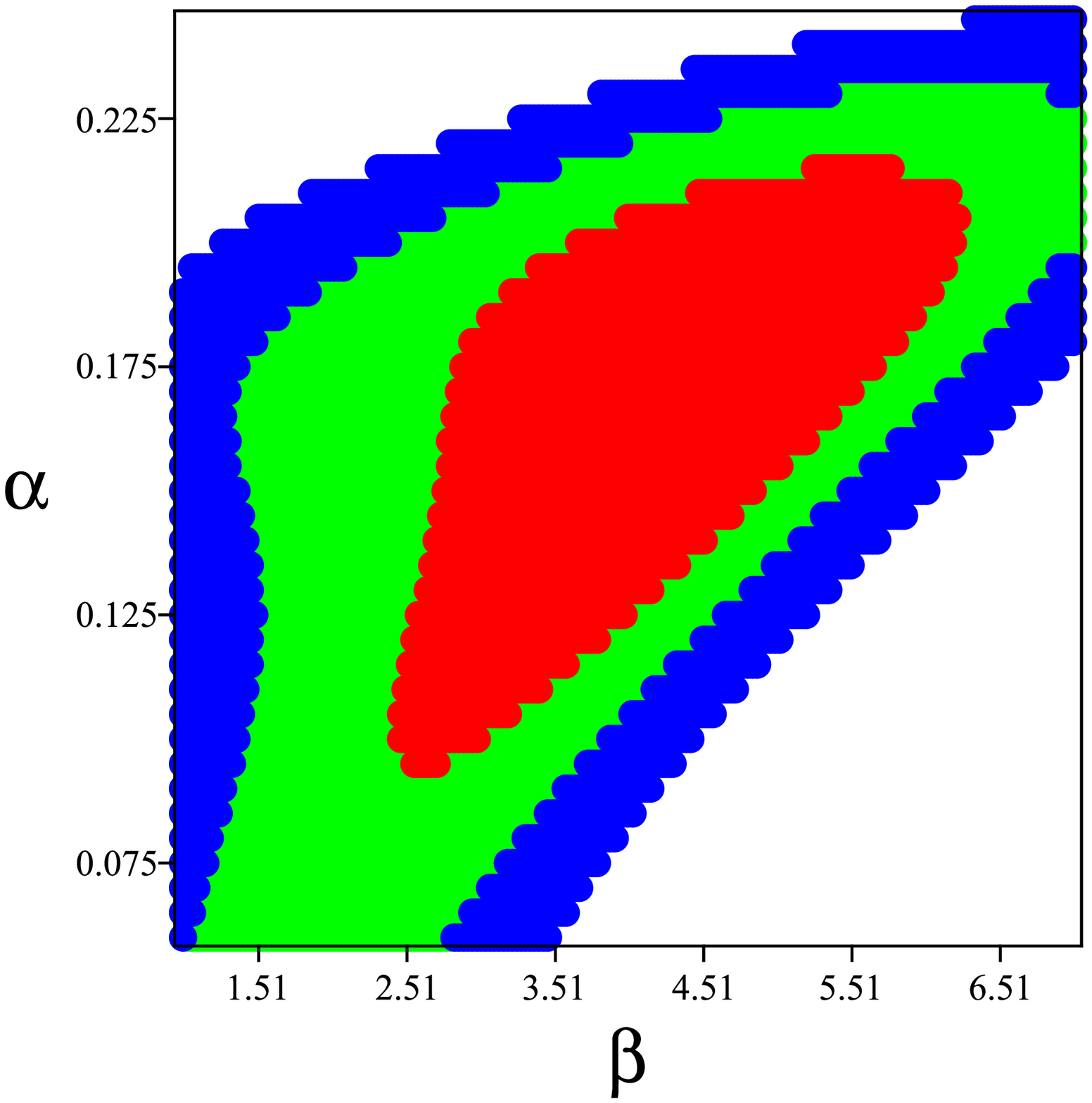}\hspace{0.1 cm}\\
Fig. 1:  The best-fitted confidence level for $\alpha$ and $\beta$ for \\ left)the power law potential and right) exponential potential\\
\end{figure}

Also, the best fitted distance modulus, $\mu(z)$, with the experimental data in both exponential and power law potential cases illustrated in Fig. 2.

\begin{figure}[t]
\includegraphics[scale=.25]{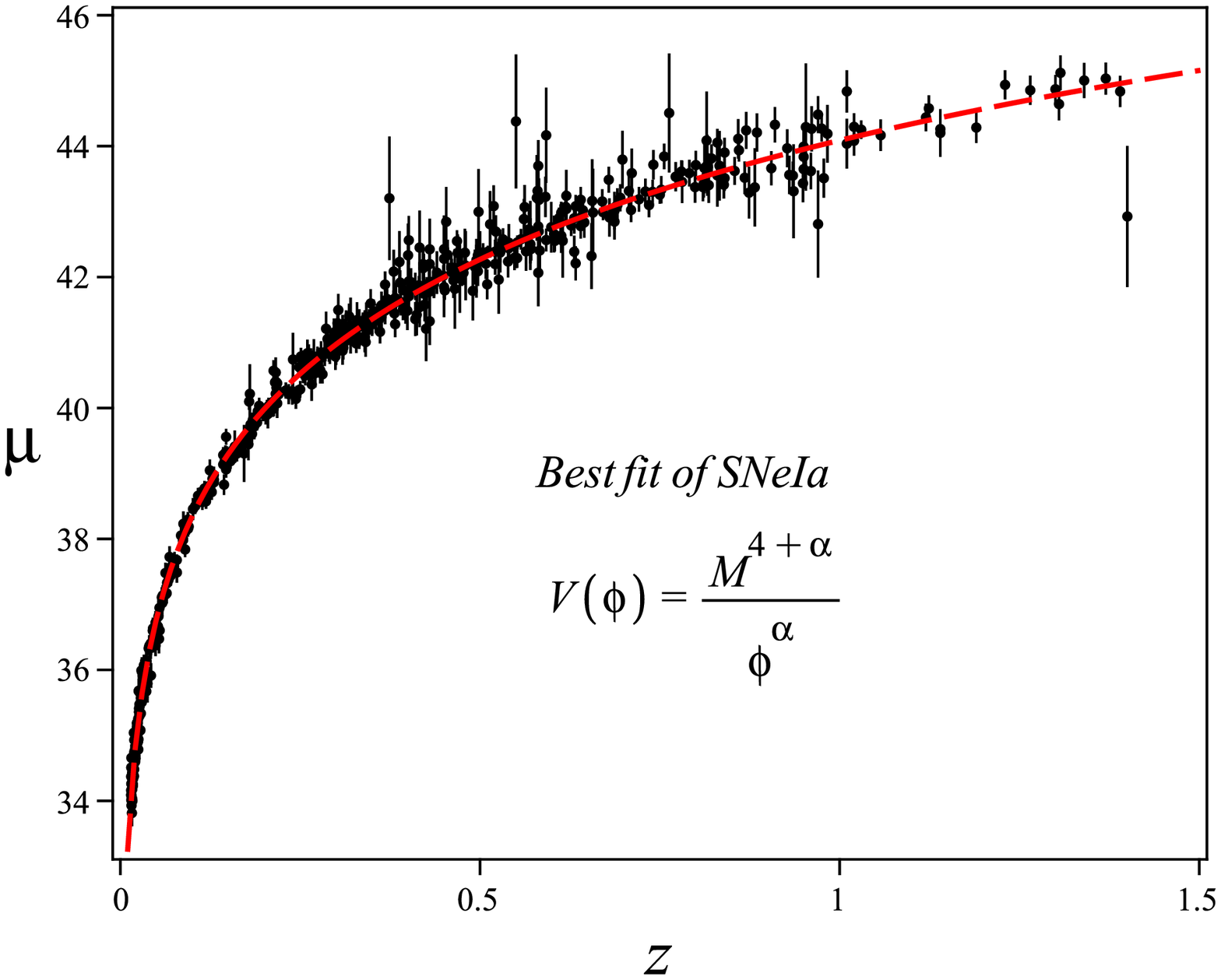}\hspace{0.1 cm}\includegraphics[scale=.25]{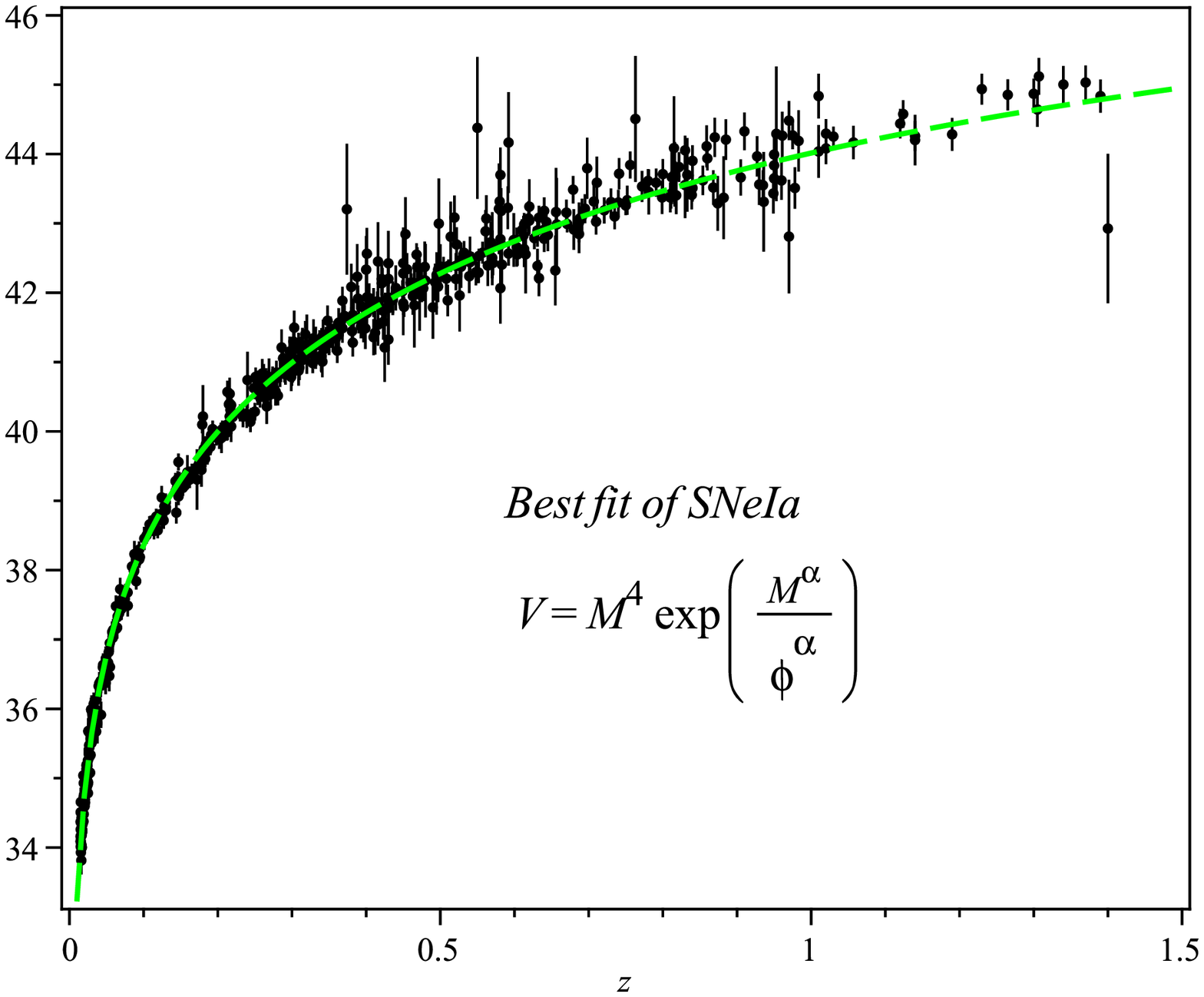}\hspace{0.1 cm}\\
Fig. 2:  The best fitted distance modulus $\mu(z)$ in both cases plotted as function of redshift \\
\end{figure}

With the best fitted model parameters, in the next section, we study in more details the chameleon mechanism by performing cosmological tests.

\section{Chameleon mechanism and cosmological test}

Complementary to the conventional chameleon mechanism given in \cite{Khoury}, here, we study the mechanism with respect to cosmological scales in different epoches. Just as an example, we take two values for the matter energy density, $\rho_m=1$ and $\rho_m=3$ where the corresponding size of the universe is $a=a_0=1$ (today) and $a=0.69$ respectively. With the best fitted model parameters, in case of power law potential, the effective potential are shown in Fig. 3: top and bottom. From the graph, as matter energy density, $\rho_m$, decreases or alternatively $a\rightarrow a_0$, the minimum effective potential shifts to larger values of $\phi_{min}$. Moreover, the steepness of the effective potential, $V_{eff}$, near the minimum, also depends on $\rho_m$; a shallow minimum corresponds to a low chameleon mass. The mass of chameleon field, $m$, increases with $\rho_m$. In other words, chameleon field has a greater contribution, as the universe approaches its earlier epoches. The same argument applies to the case of exponential potential case (Fig. 4).

\begin{figure}[t]
\includegraphics[scale=.25]{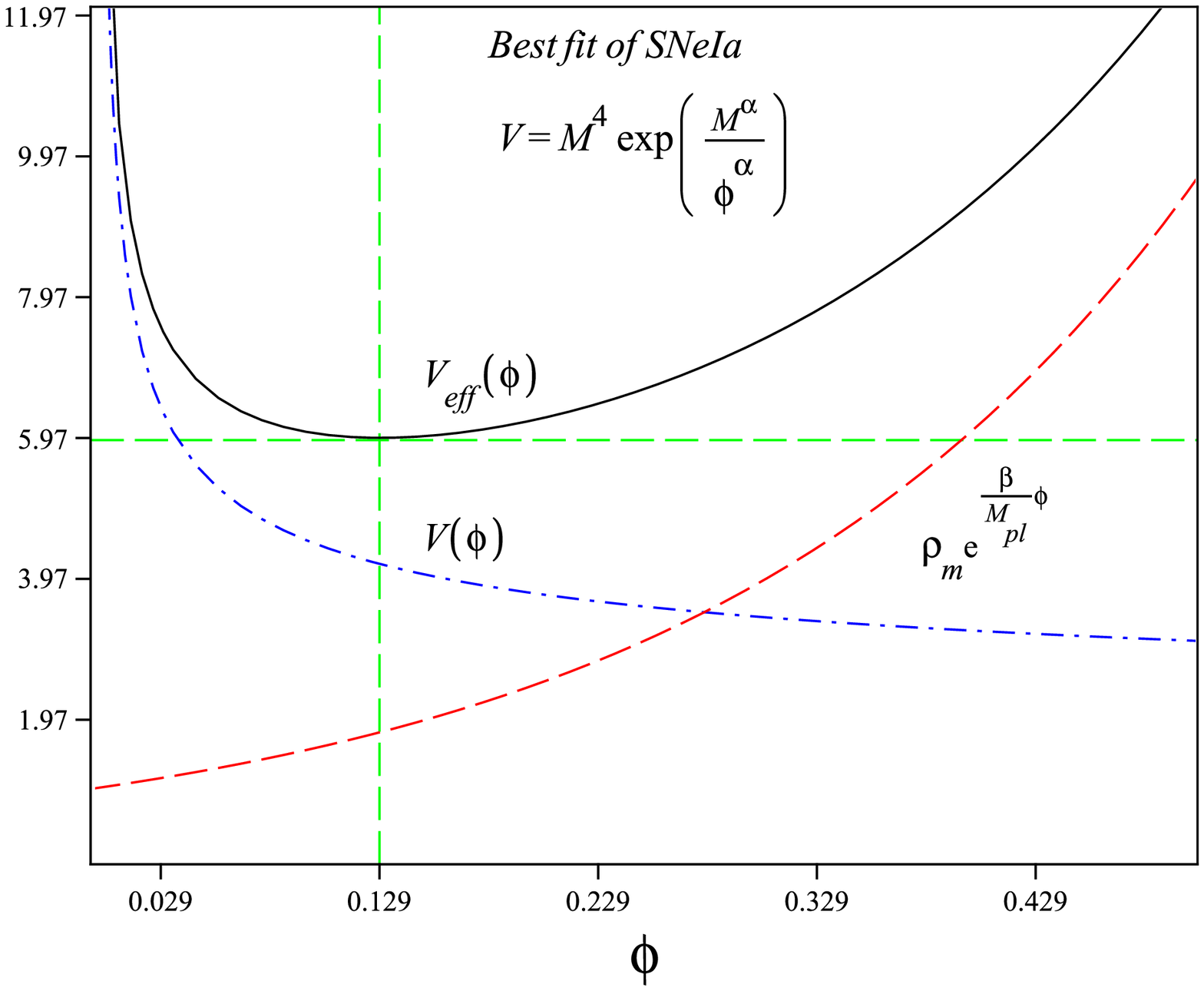}\hspace{0.1 cm}\includegraphics[scale=.25]{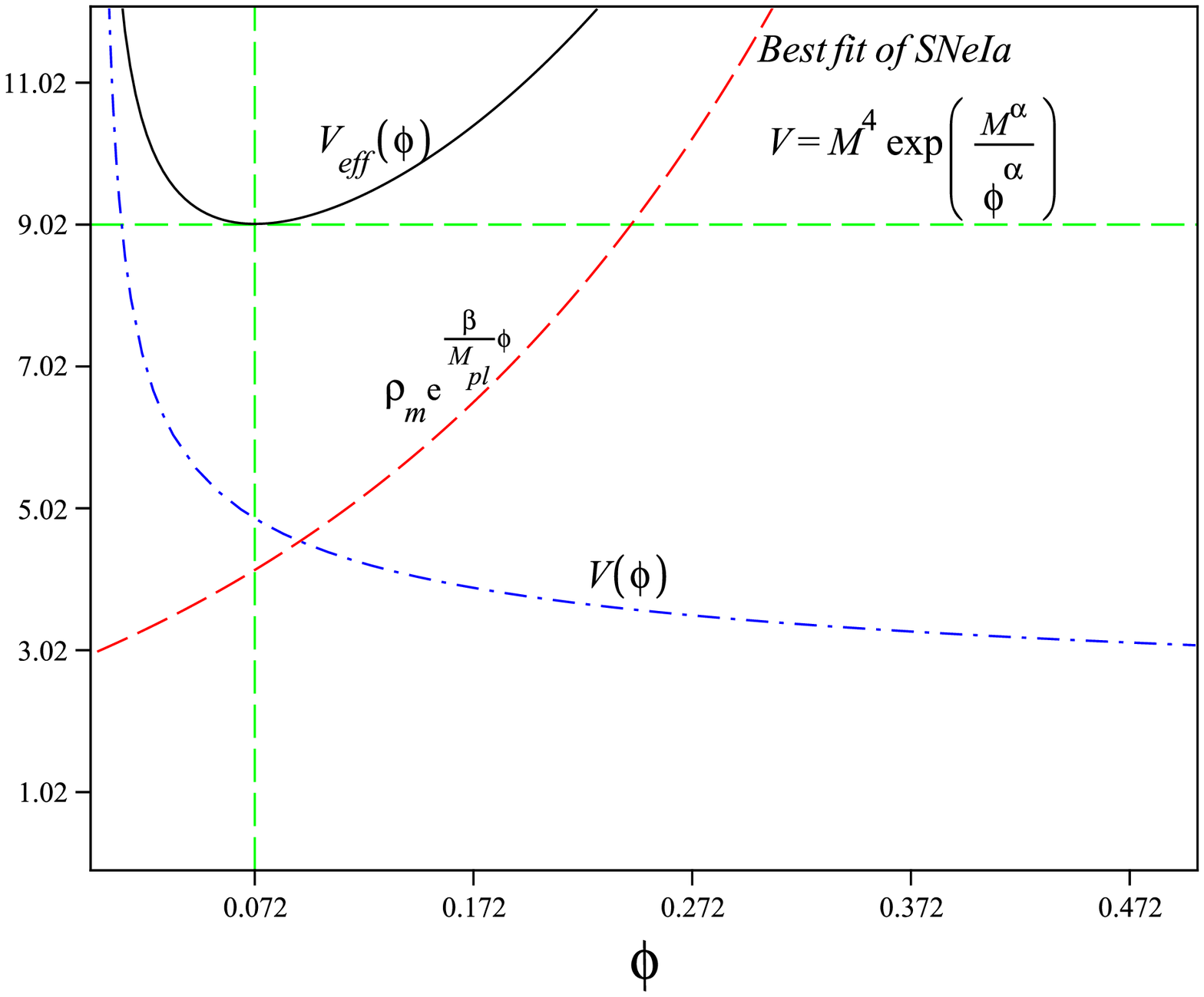}\hspace{0.1 cm}\\
Fig.3:  The best fitted behavior of chameleon effective potential, $V_{eff}(\phi)$, in case of exponential potential for (Top)$\rho_m=1$ and (Bottom)$\rho_m=3$. \\
\end{figure}
\begin{figure}[t]
\includegraphics[scale=.25]{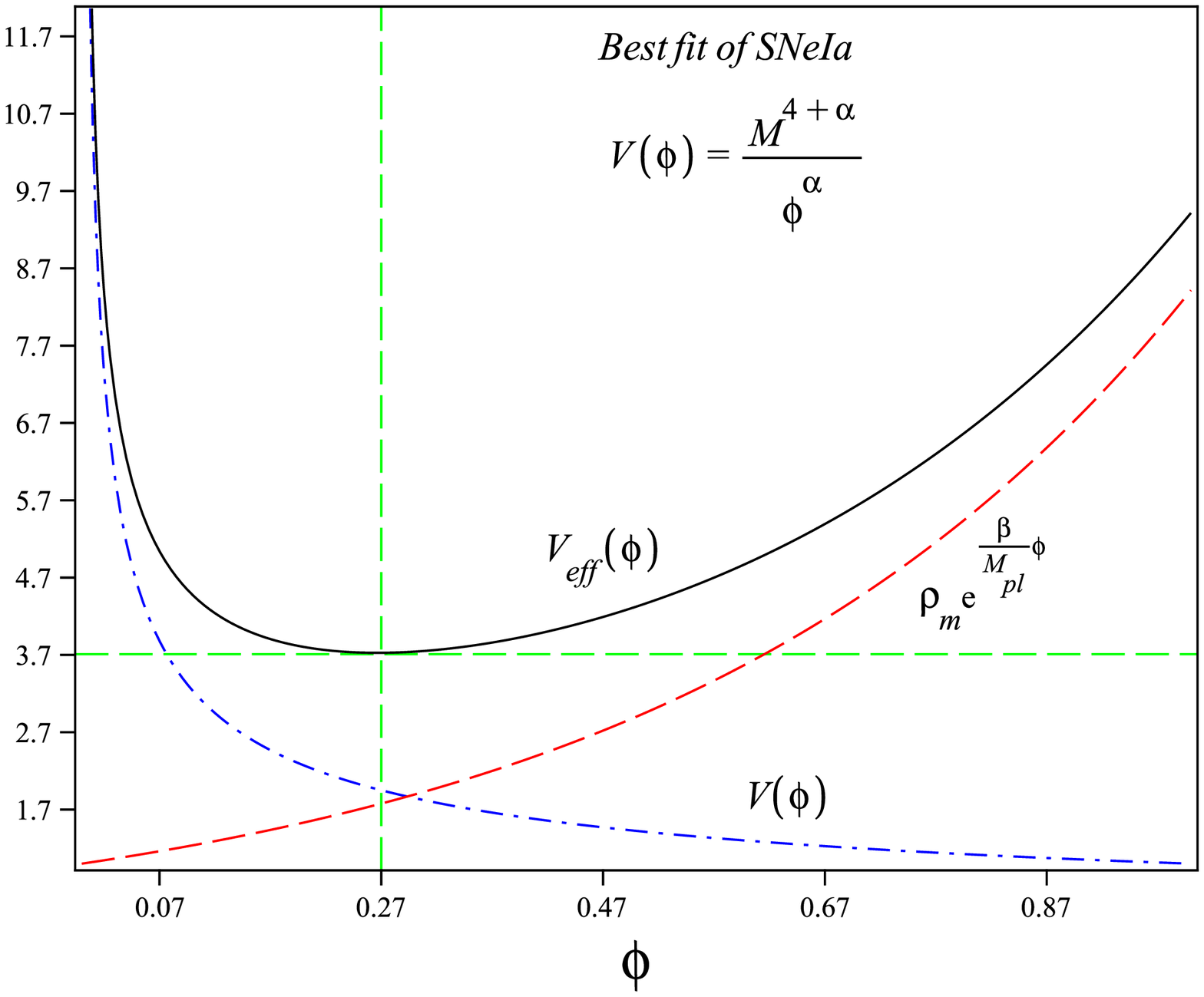}\hspace{0.1 cm}\includegraphics[scale=.25]{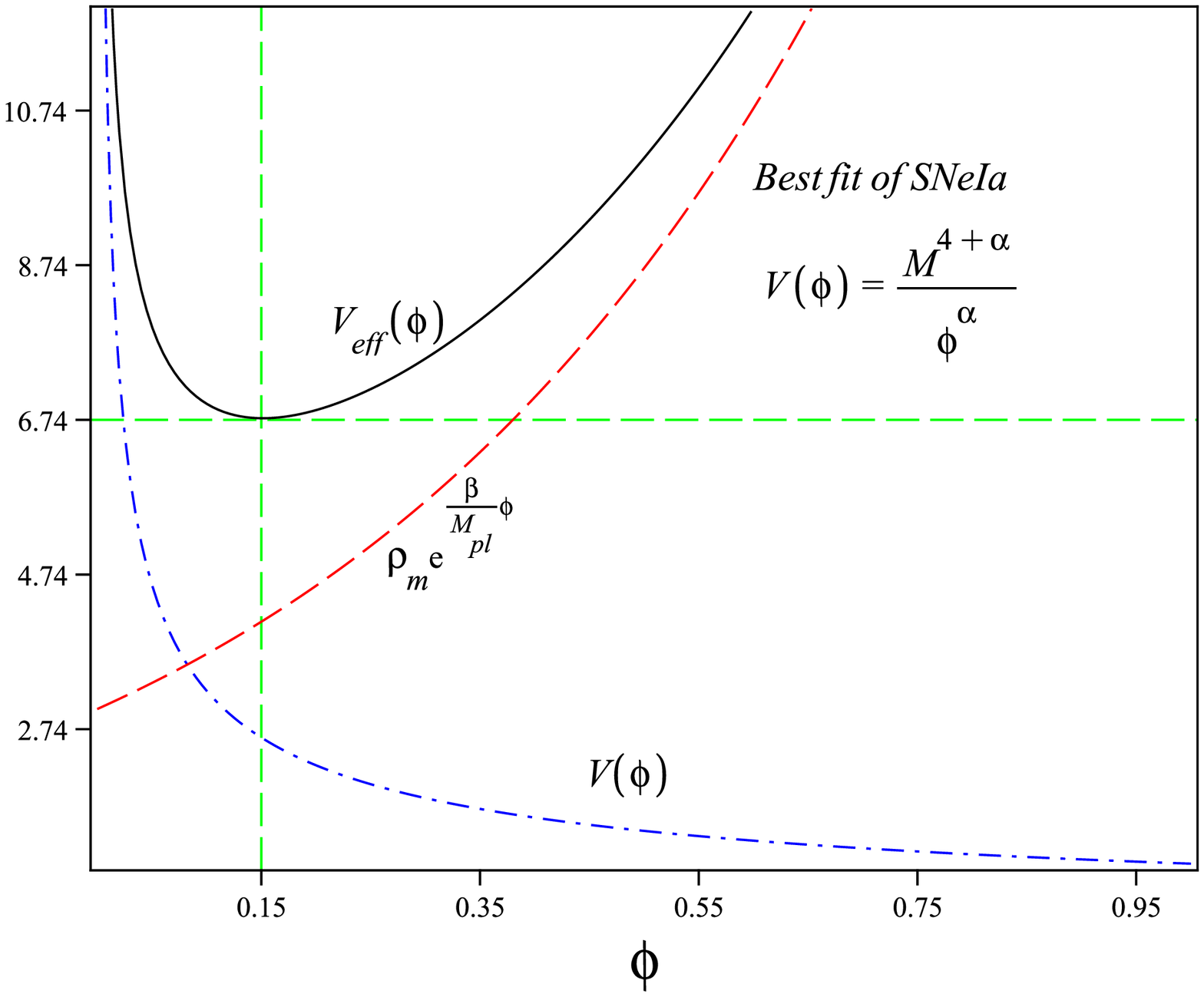}\hspace{0.1 cm}\\
Fig. 4: The best fitted behavior of chameleon effective potential, $V_{eff}(\phi)$, in case of power law potential for (Top)$\rho_m=1$ and (Bottom)$\rho_m=3$. \\
\end{figure}

A cosmological quantity that usually considered as a significant parameter to measure the reliability and affectiveness of cosmological models is equation of state (EoS) parameter. In Fig. 5, a comparison between power law and exponential potential scenarios for the best fitted model parameters with observation is shown. The graph illustrates that while, up to about $z\simeq 1.5$, the trajectories of effective EoS parameter in both scenarios overlap each other, they completely resolved afterward. Since there is no direct measurement of EoS parameter for the universe, one can not decide which form of potential is observationally privileged. The graph also shows that the current effective EoS parameter for the best fitted model parameters in both cases is $\omega_{eff} \simeq -0.46$.
Next, we examine our models with cosmological data for Hubble parameter.

\begin{figure}[t]
\includegraphics[scale=.25]{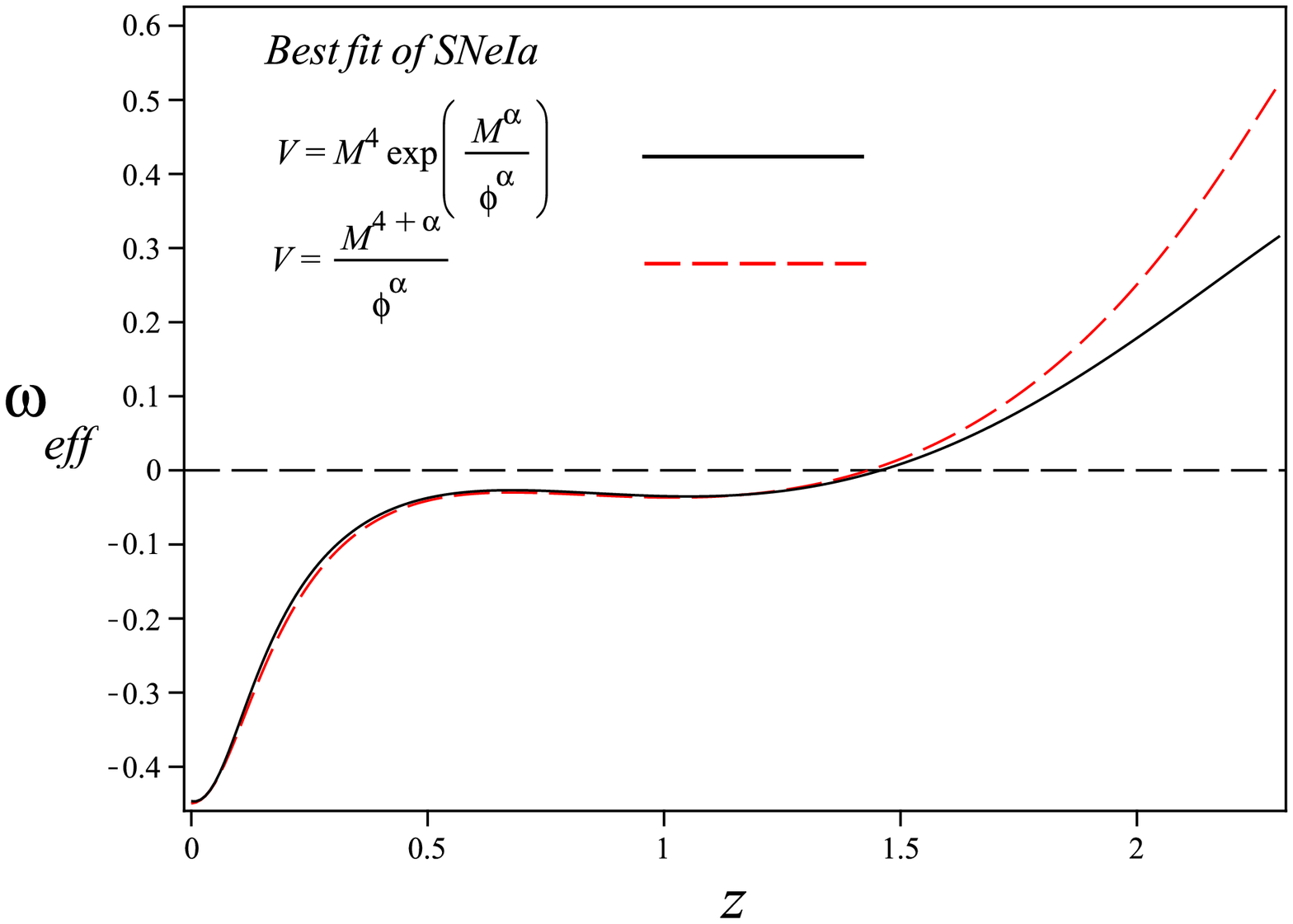}\hspace{0.1 cm}\\
Fig. 5:  The best fitted effectivce EoS parameter for power law and exponential potential. \\
\end{figure}

In Fig. 6 the best fitted Hubble parameter for both potential functions obtained from numerical calculation is compared with the observational data. As can be seen, for the redshift $z\leq0.5$ the two curves overlap and relatively agree with the observational data. Within the distance $0.5\leq z\leq 1.5$, they again reasonably fit the data whereas have quite distinct shape. Nevertheless, similar to the case of EoS parameter, the two curves begin to resolve at $z\geq 1.5$. Again, the two models satisfy the observational data within the observationally fitted interval $0\leq z\leq 1.5$. However, at this stage, since the model is not tested against data for higher redshifts, $z\geq2$, one still can not decide which of the two scenarios is observationally preferable.

\begin{figure}[t]
\includegraphics[scale=.25]{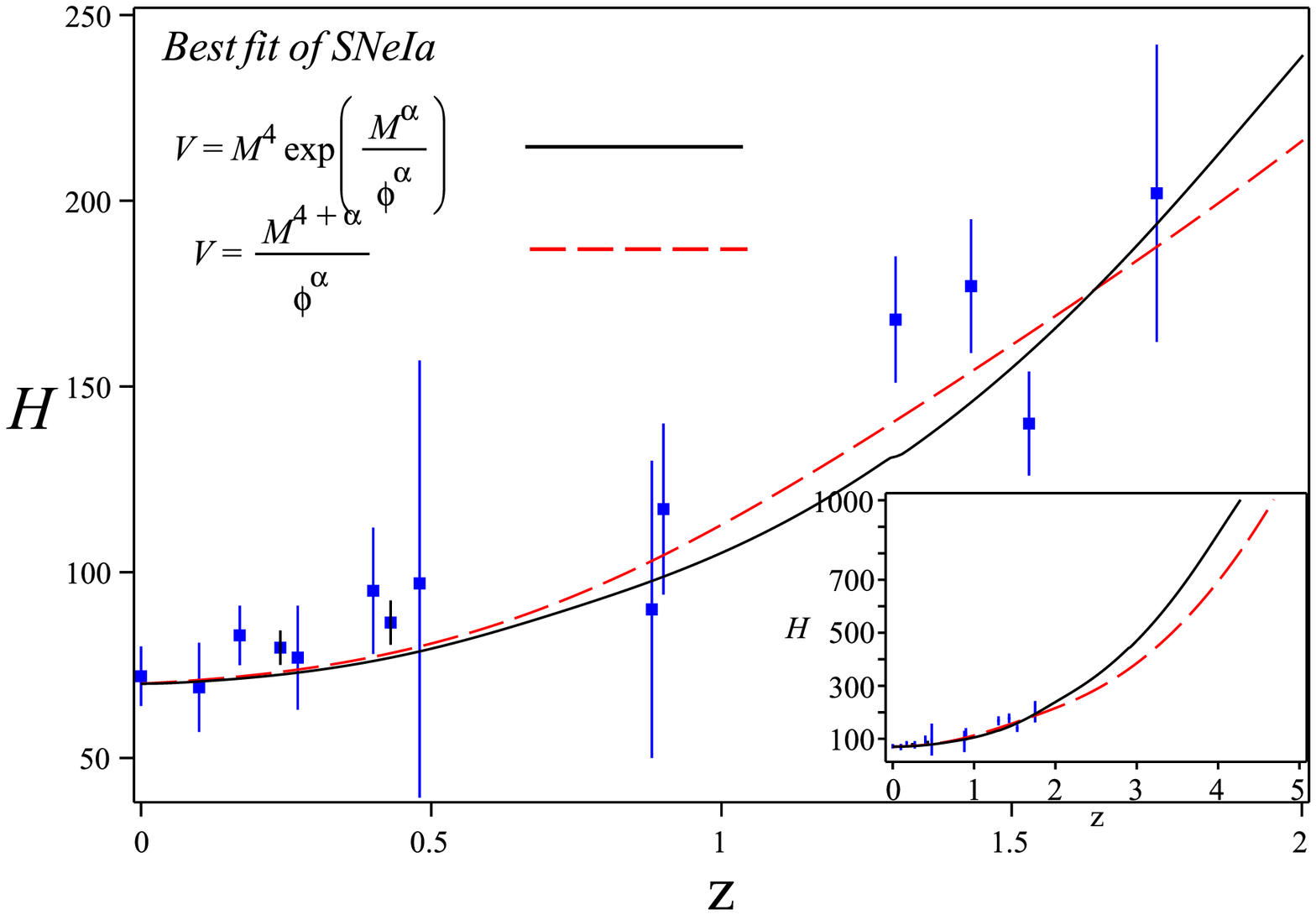}\hspace{0.1 cm}\\
Fig. 6:  The best fitted Hubble parameter for power law and exponential potential against observational data. \\
\end{figure}

Next, we choose to compare the two models against observational data for velocity drift ~\cite{Liske} where the data distributed over $1.5\leq z\leq 5$. With the velocity drift given by
\begin{eqnarray}\label{vdrift}
\dot{v}=cH_0-\frac{cH(z)}{1+z},
\end{eqnarray}
a comparison of the two models with the observational data for high redshift are shown in Fig. 7.

\begin{figure}[t]
\includegraphics[scale=.25]{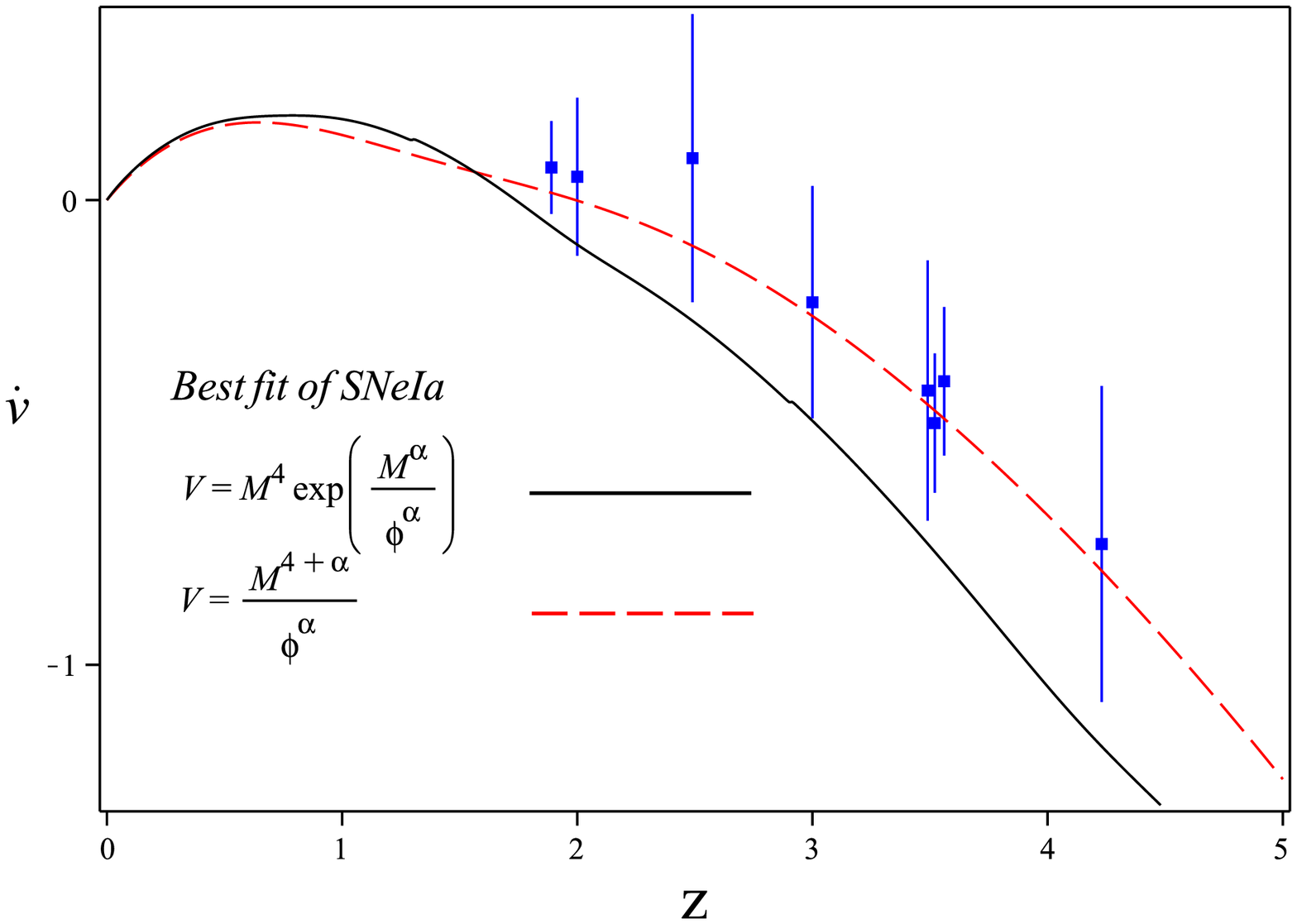}\hspace{0.1 cm}\\
Fig. 7:   The graph of best fitted velocity drift $\dot{v}$ for exponential and \\ power law potentials in comparison with the data
\end{figure}

As can be seen from the graph one can choose power law potential function a more natural choice over exponential potential function. This is also in agreement with the discussion given in \cite{khori2}.

\section{Summary and remarks}

This paper examines the chameleon mechanism in cosmological scales by taking into consideration power-law and exponential potential functions. The work is supported by best fitting the model parameters with the observational data. Departing from conventional approach to chameleon mechanism, we revisit the model of interacting chameleon field in the framework of cosmological scales. The result shows that in both cases of power-law and exponential potential the mass of interacting chameleon field reduces with the matter energy density. The two scenarios of power-law and exponential potential functions directly and indirectly are tested against observational data. We find that in the range of $0<z\lesssim 1.5$, the dynamic of the effective EoS parameter is independent of the underlined models. However the circumstance is different in higher redshifts, i.e. $z\geq 1.5$. The EoS parameter also shows that the universe begins to accelerate at about $z\simeq 0.2$ and currently is in quintessence era. We perform two observational tests comparing the dynamics of the model in two scenarios. First, the best fitted Hubble parameter derived from numerical computation in both cases are compared with the observational data. While both models are relativity in good agreement with the data for $z\geq 1.5$, no comparison can be made with the data for high redshifts. Second, we compare the best fitted velocity drift computed in both scenarios with the data for high redshift, $1.5\leq z \leq 4.5$. The test shows a better agreement between the data and best fitted velocity drift in power law potential case.

\end{document}